\documentclass[10pt,conference]{IEEEtran}
\usepackage[english]{babel}
\usepackage[utf8x]{inputenc}

\usepackage{microtype}

\usepackage{graphicx}
\graphicspath{{./img/}}
\usepackage{subcaption}

\usepackage{booktabs}
\usepackage{listings}

\usepackage[bookmarks=false]{hyperref} 

\usepackage[cmex10]{amsmath} 
\usepackage{cite}
\usepackage{balance}

\usepackage[top=0.75in,bottom=1in,left=0.63in,right=0.65in]{geometry}

\usepackage[absolute,showboxes]{textpos}

\makeatletter
\def\ps@headings{\def\@oddhead{\mbox{}\scriptsize\rightmark \hfil \thepage}\def\@evenhead{\scriptsize\thepage \hfil \leftmark\mbox{}}\def\@oddfoot{}\def\@evenfoot{}}
\makeatother
\title{Uncovering Vulnerable Industrial Control Systems from the Internet Core}

\author{
  
  \IEEEauthorblockN{Marcin Nawrocki}
  \IEEEauthorblockA{Freie Universit\"at Berlin\\marcin.nawrocki@fu-berlin.de}

 \and

  \IEEEauthorblockN{Thomas C. Schmidt}
  \IEEEauthorblockA{HAW Hamburg\\t.schmidt@haw-hamburg.de}

 \and

  \IEEEauthorblockN{Matthias W\"ahlisch}
  \IEEEauthorblockA{Freie Universit\"at Berlin\\m.waehlisch@fu-berlin.de}}

\usepackage{xargs}
\usepackage[pdftex,dvipsnames]{xcolor}
\usepackage[colorinlistoftodos,prependcaption,textsize=tiny]{todonotes}
\newcommandx{\unsure}[2][1=]{\todo[linecolor=red,backgroundcolor=red!25,bordercolor=red,#1]{#2}}
\newcommandx{\change}[2][1=]{\todo[linecolor=blue,backgroundcolor=blue!25,bordercolor=blue,#1]{#2}}
\newcommandx{\info}[2][1=]{\todo[linecolor=OliveGreen,backgroundcolor=OliveGreen!25,bordercolor=OliveGreen,#1]{#2}}
\newcommandx{\improvement}[2][1=]{\todo[linecolor=Plum,backgroundcolor=Plum!25,bordercolor=Plum,#1]{#2}}
\newcommandx{\done}[2][1=]{\todo[linecolor=Gray,backgroundcolor=White!25,bordercolor=Gray,#1]{#2}}

\usepackage{pifont}
\newcommand{\cmark}{\ding{51}}\newcommand{\xmark}{\ding{56}}\newcommand{\onec}{\ding{202}}\newcommand{\twoc}{\ding{203}}\newcommand{\threec}{\ding{204}}

\usepackage{xspace}
\newcommand{\eg}{\textit{e.g.,}~}
\newcommand{\ie}{\textit{i.e.,}~}
\newcommand{\etc}{\textit{etc.}~}
\newcommand{\one}{({\em i})\xspace}
\newcommand{\two}{({\em ii})\xspace}
\newcommand{\three}{({\em iii})\xspace}
\newcommand{\four}{({\em iv})\xspace}

\makeatletter
\renewcommand{\paragraph}[1]{\noindent{\bf #1.}\hspace{0.25ex \@plus1ex \@minus.2ex}}
\makeatother

\let\orgautoref\autoref
\renewcommand{\autoref}
{\def\sectionautorefname{Section}\def\subsectionautorefname{Section}\def\subsubsectionautorefname{Subsection}\orgautoref}

\IEEEoverridecommandlockouts\IEEEpubid{\makebox[\columnwidth]{978-1-7281-4973-8/20/\$31.00 ~\copyright~2020 IEEE \hfill} \hspace{\columnsep}\makebox[\columnwidth]{ }}

\begin{document}
\bstctlcite{IEEEexample:BSTcontrol}

\setlength{\TPHorizModule}{\paperwidth}
\setlength{\TPVertModule}{\paperheight}
\TPMargin{5pt}
\begin{textblock}{0.8}(0.1,0.02)
	\noindent
	\footnotesize
	\centering
	If you cite this paper, please use the NOMS reference:
	M. Nawrocki, T. C. Schmidt, M. Wählisch.~2020.
	Uncovering Vulnerable Industrial Control Systems from the Internet Core
	\emph{Proceedings of 17th IEEE/IFIP Network Operations and Management Symposium (NOMS)}, 2020.
\end{textblock}

\maketitle

\begin{abstract}
Industrial control systems (ICS) are managed remotely with the help of dedicated protocols that were originally designed to work in walled gardens. Many of these protocols have been adapted to Internet transport and support  wide-area communication.  
ICS now exchange insecure traffic on an inter-domain level, putting at risk not only common critical infrastructure, but also the Internet ecosystem (\eg DRDoS~attacks).

In this paper, we uncover \emph{unprotected} inter-domain ICS traffic at two central Internet vantage points, an IXP and an ISP.
This traffic analysis is correlated with data from honeypots and Internet-wide scans to separate industrial from non-industrial ICS traffic.
We provide an in-depth view on Internet-wide ICS communication. Our results can be used
\one to create precise filters for potentially harmful non-industrial ICS traffic, and
\two to detect ICS sending unprotected inter-domain ICS traffic, being vulnerable to eavesdropping and traffic manipulation attacks.
\end{abstract}

\section{Introduction}

Industrial control systems (ICS) are used to monitor and control industrial environments.
Deployments can range from a few controllers in a factory to large distributed systems that monitor critical infrastructures.
The underlying ICS communication is based on specialized, often proprietary protocols.

\begin{figure}
\centering
  \centering
  \includegraphics[width=0.95\linewidth]{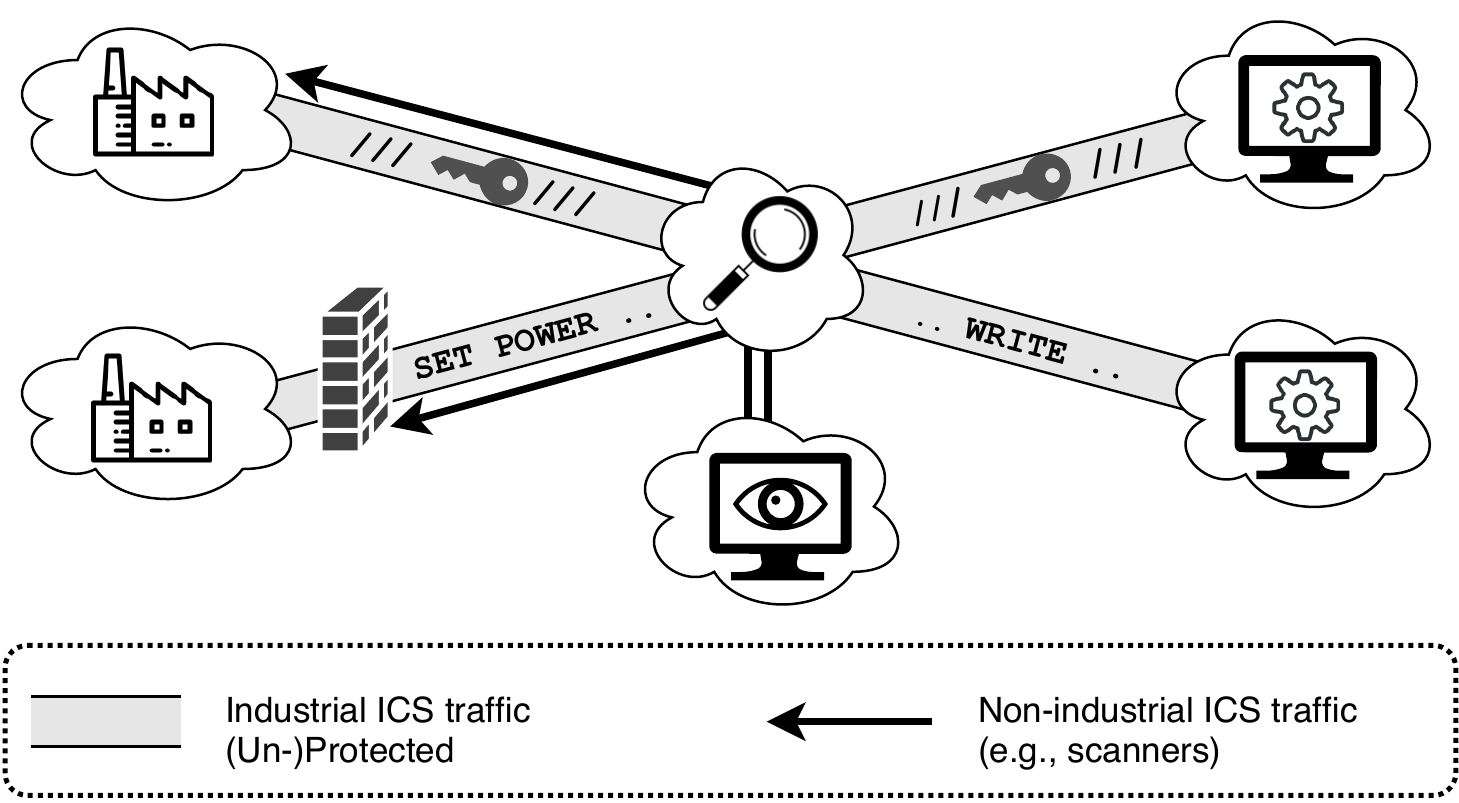}
  \caption{Analyzing unprotected ICS protocols.}
  \label{fig:unprotectedICStraffic}
\end{figure}

Originally, ICS protocols were designed to operate in closed environments, which do not require authentication and encryption.
The lack of security features in ICS protocols remained largely unnoticed due to the deployment in isolated (trusted) environments.
This changed recently when ICS protocols have been stacked onto IP, enabling the management of ICS controllers via the global Internet.
Such communication requires protective measures, either via secure tunnels between trusted domains or end-to-end authentication and encryption.
Visible (unencrypted) ICS traffic is particularly dangerous since it is prone to eavesdropping and manipulation attacks.
Traffic traces also hint attackers to potentially open ICS services without the need to perform suspicious scans.
\autoref{fig:unprotectedICStraffic} sketches encrypted and visible traffic flows between ICS.
It also shows a passive vantage point and an active scanner, which might be blocked by a firewall.
Note that the firewall does not help in the case of man-in-the-middle manipulation attacks.

In this paper, we provide the first comprehensive analysis of the visibility of \emph{unprotected} ICS traffic across network domains.
In contrast to previous work \cite{bodenheim2014impact, 7906943} which reveals reachable ICS services, we explore the communication of the whole ICS ecosystem, from the ICS controllers to the management stations.
We show that ICS systems are controlled remotely without any protective mechanisms, harming both the Internet as well as the industrial infrastructure.
Our results attract attention to the insecure usage of ICS protocols and motivate secure ICS deployments with encrypted tunnels.
In detail, our contributions are the following.
\begin{enumerate}
  \item We present the first analysis of inter-domain ICS traffic at two central Internet vantage points, an Internet Exchange Point and an Internet Service Provider, covering 6 months.
  \item We find new unprotected ICS deployments which are undetected by recent scan projects.
  \item We classify industrial and non-industrial ICS traffic based on cross-correlations with other data sources such as honeypots.
  \item  We assess common tools for implementing our proposed methodology to allow for future long-term monitoring and mitigation.
\end{enumerate}

The remainder of this paper is structured as follows.
Section~\ref{sec:background} presents related work about ICS protocols.
Section~\ref{sec:analysis-identification} introduces our methodology and data sources to identify ICS traffic.
Section~\ref{sec:analysis-properties} presents basic properties of ICS traffic seen at the IXP and ISP.
Section~\ref{sec:analysis-classification} proposes a method to separate industrial and non-industrial ICS traffic.
Section~\ref{sec:trafficFeatures} analyzes industrial ICS traffic in detail.
Section~\ref{sec:conclusion} concludes our findings.

\section{Background and Related Work}
\label{sec:background}

\subsection{A Glimpse into ICS Protocol Security}

\begin{table*}  \centering
\caption{Overview of ICS Protocols. [ND/HD: Normal/Heuristic Dissector, C: Censys, S: Shodan, R: Rapid7, K: Kudelski]
\label{tab:ics-overview}}
  \begin{tabular}{ l l l c c c c r }
  
  \toprule

  & & 										& Wirshark 	& Min. \#~Bytes to & Scan 	& Honeypot & \#  \\
  Standard / Protocol  		& Ports 		& Use Case		& Dissectors 	& identify protocol 	& Projects 	& Software 	& CVEs \\ \midrule
  Modbus 			& 502 			& Process automation		& ND 		& 74 B 		& C/S/K 	& \cmark 	& 23  \\  
  Siemens S7 			& 102 			& Process automation		& HD 		& 93 B 		& C/S 		& \cmark 	& 7 \\
  EthernetIP 			& 2221, 2222, 44818	& Process automation		& ND 		& 74 B 		& S 		& \cmark 	& 9 \\  
  BACnet 			& 47808-47823 		& Building management	& ND 		& 46 B 		& C/S/R/K 	& \cmark 	& 7 \\
  
  DNP3 				& 20000 		& Smart grids		& ND/HD 	& 62 B 		& C/S 		& \xmark 	& 39 \\  
  HART IP 			& 5094 			& Process automation		& ND 		& 78 B 		& S 		& \xmark 	& 6 \\
  IEC60870-5-104 		& 2404 			& Smart grids		& ND 		& 76 B 		& S 		& (\xmark) 	& 0 \\
  ANSI C12.22 			& 1153 			& Metering		& ND 		& n/a 		& \xmark 	& \xmark 	& 0 \\
  OMRON FINS 			& 9600 			& Process automation		& ND 		& 54 B 		& S 		& \xmark 	& 7 \\  
  IEC61850 (mms)		& 102 			& Smart grids		& ND/HD 	& 144 B 	& \xmark 	& \xmark 	& 0 \\

  \midrule

  Codesys 			& 2455 			& Smart grids		& \xmark & & S & \xmark & 20 \\
  GE-SRTP 			& 18245, 18246 		& Process automation		& \xmark & & S & \xmark & 7 \\  
  Niagara Fox 			& 1911, 4911 		& Building management 	& \xmark  & & C/S/K & \xmark  & 5 \\
  MELSEC-Q 			& 5006, 5007 		& Process automation		& \xmark & & S & \xmark & 2 \\
  ProConOS 			& 20547 		& Process automation		& \xmark & & S & \xmark & 1 \\
  PCWorx 			& 1926 			& Process automation		& \xmark & & S & \xmark & 0 \\
  Crimson 			& 789 			& Process automation		& \xmark & & S & \xmark & 0 \\
  
  ICCP-TASE.2 			& 102 			& Smart grids		& \xmark & & \xmark & \xmark & 8 \\
  
  \bottomrule

\end{tabular}

\end{table*}

ICS protocols are deployed in four major application areas \cite{7906943}:
\one process automation, \two building management, \three smart grids including power plants, and \four metering infrastructures, an overview is presented in \autoref{tab:ics-overview}.
All of these scenarios require security support when the ICS devices are interconnected via untrusted networks.

Vulnerable ICS deployments have been highlighted since several years \cite{meixell2013out, klick2015internet}. 
The first reported incident is an unauthorized manipulation of an ICS which led to a pipeline explosion back in 1982 \cite{miller2012survey}.
Although the absolute number of reported ICS incidents is fairly low \cite{miller2012survey}, a single incident can be hazardous.
To understand and improve the protection of ICS deployments, multiple efforts have been undertaken, including \one the development of honeypots, \two Internet-wide scans to find open ICS devices,  \three the improvement of intrusion detection systems for ICS, and \four the modeling and surveying of the ICS ecosystem.

ICS specific honeypots have been developed \cite{vasilomanolakis2015did, wilhoit2013s, winn2015constructing, bernieri2019mimepot} to understand the origin, frequency, and sophistication of attacks on ICS services.
ICS services are popular victims. ICS honeypots receive significantly more requests after being listed on public scanning sites such as Shodan \cite{serbanescu2015ics}.

Two well-known scan projects, Censys and Shodan, detect globally reachable ICS services \cite{bodenheim2014impact, 7906943}.
Such scan results can be used to asses the security of ICS in individual countries\cite{ceron2019online}.
ICS scans are dominated by few recurrent scanners \cite{ferretti2019characterizing} and captured within few days by honeypot deployments \cite{bodenheim2014impact}.
Mirian \emph{et al.}~\cite{7906943} measured the increase of open ICS services of up to 20~\% in 4~months.

Dedicated intrusion detection systems (\eg for smart meters~\cite{berthier2011specification}) and extensions to common IDS tools (\eg Snort and Bro \cite{morris2012retrofit, lin2013adapting, bajtovs2020multi}) have been proposed.
Valdes \cite{valdes2009intrusion} introduces an architecture that monitors ICS traffic for irregular patterns.
Taking into account recent, distributed ICS deployments, Zhang \cite{zhang2011distributed} proposed a distributed multi-layered system.

ICS traffic patterns have been compared with SNMP traffic \cite{barbosa2012first}.
Both, ICS protocols and SNMP, show stable, periodical traffic patterns with a small number of constant host changes.
However, ICS traffic does not present diurnal patters or self-similar correlations, features known from traditional network protocols \cite{barbosa2012difficulties}.
In contrast to our approach, the data for this comparison was collected directly at the corresponding edge-network (traditional network, ICS-facilities).
So no protocol classification was necessary.

ICS have been surveyed in several publications introducing historical background, taxonomies, and current security vulnerabilities \cite{igure2006security, liu2012cyber, ralston2007cyber, zhu2010scada, 6142258, rubio2019current}.
The number of Common Vulnerabilities and Exposures (CVEs) for ICS implementations grows steadily.
Vulnerabilities are often discovered by simple fuzzing techniques \cite{shapiro2011identifying, devarajan2012unraveling}.
Also, the ICS ecosystem requires a secure supply chain\cite{hou2019understanding}.
Recent studies show the high DDoS potential of BACnet by analysing IXP and ISP packet samples over a period of 48~hours~\cite{gasser2017amplification}.
Yet, still open is a longitudinal analysis of unprotected ICS communication deployed in the global Internet.

\subsection{The Problem of Unprotected ICS Protocols}

Most of the common ICS protocols lack protection by design and are susceptible to eavesdropping and traffic manipulation attacks.
The only exception is Niagara Fox, which provides authentication.
However, authentication alone is insufficient.
Attackers can scout their target and prepare a targeted attack without communicating with the ICS devices at all.
Recent malware \cite{talosVpnfilter} exploits passive recording of ICS traffic traversing small enterprise routers.
Such eavesdropping of unprotected ICS traffic is also possible on the inter-domain level.

Furthermore, it is important to note that infrastructure-based protections such as firewalls or NAT only partially help.
They may prevent discovering ICS devices by \emph{active} scanning but do not protect against \emph{passive} listening and spoofed replay attacks.

In this paper, we analyze the highly vulnerable part of the ICS ecosystem; those cases where operators interconnect their systems without any protection.
This is challenging because unprotected industrial ICS traffic is suppressed by noise such as scan traffic.

\subsection{ICS Scans Seen from an Internet Telescope}

To motivate our aim for a detailed classification of ICS traffic, we briefly analyse data from the CAIDA/UCSD network telescope.
This data source captures backscatter traffic from randomly spoofed DDoS attacks or Internet-wide scans of the /8 CAIDA/UCSD darknet.
Any incoming traffic to the telescope is inter-domain and non-industrial.

\begin{figure}
  \centering\includegraphics[width=\linewidth]{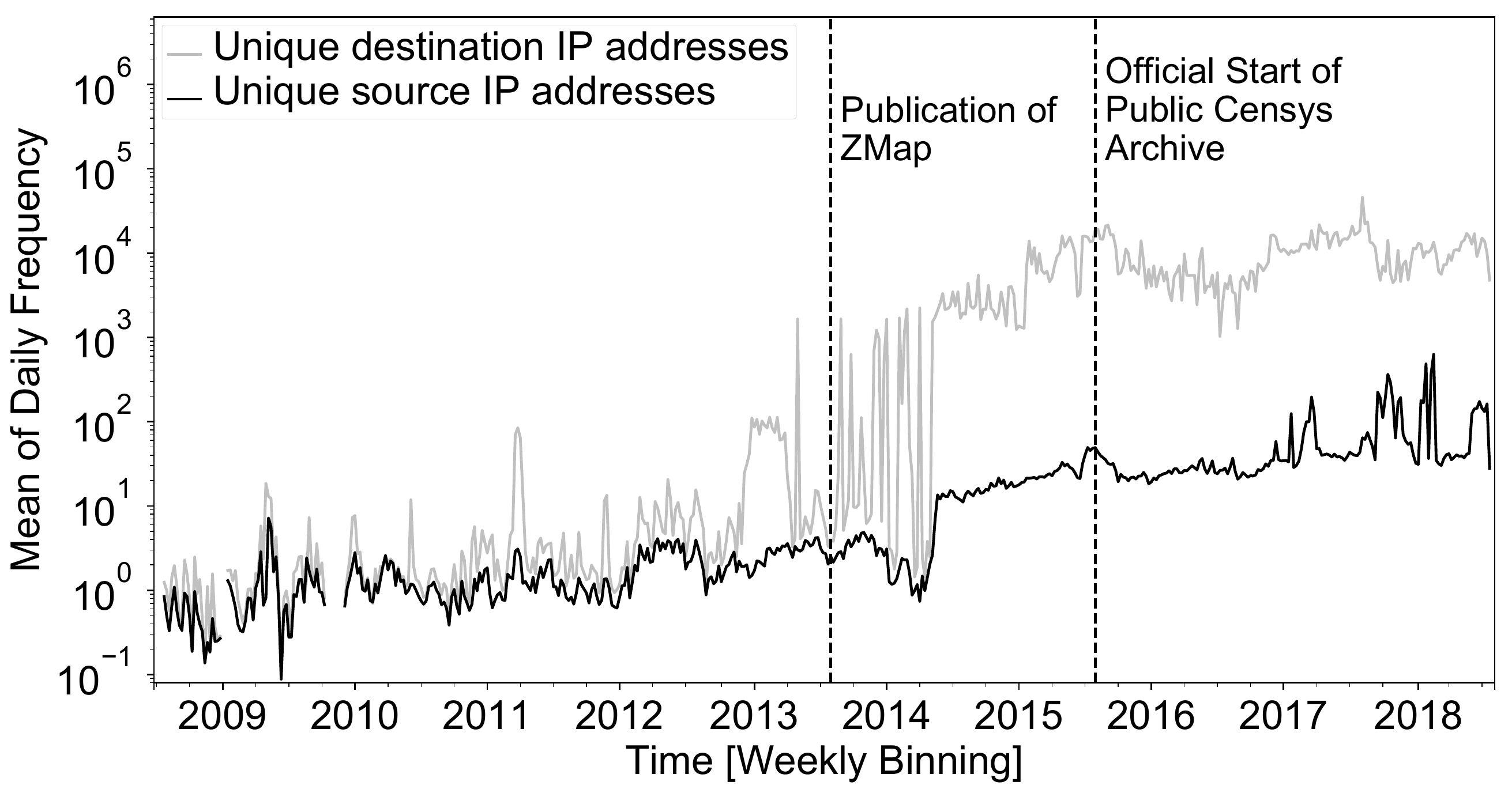}
  \caption{Internet-wide scanning of Modbus (TCP/502) observed at the CAIDA network telescope.
  We highlight research activities around one of the most common ICS scanners.}
  \label{fig:ioda_all}
\end{figure}

\autoref{fig:ioda_all} shows the daily activity for Modbus (TCP/502), measured at the telescope.
There is almost no activity visible until the beginning of 2014.
Then, the amount of destination IP addresses that received data on the Modbus port increased by three orders of magnitude.
The number of source IP addresses that sent data to the telescope increased by roughly one order of magnitude, indicating scanning from a small set of hosts.
The sudden upturn in scan activities can be explained by 
\one increased media coverage of ICS systems and 
\two increased research interest and consequently publicly available scan tools.
Our observations correlate with the start of the ZMap and Censys projects.
We saw no correlation with Shodan, which started to index ICS infrastructures in 2009 and added ICS protocols in 2012.

This brief analysis does not only highlights the increasing interest in ICS protocols but also the need for a careful methodology to analyse ICS traffic.

\begin{figure*}
  \center
  \begin{subfigure}{.32\textwidth}
    \centering\includegraphics[width=\linewidth]{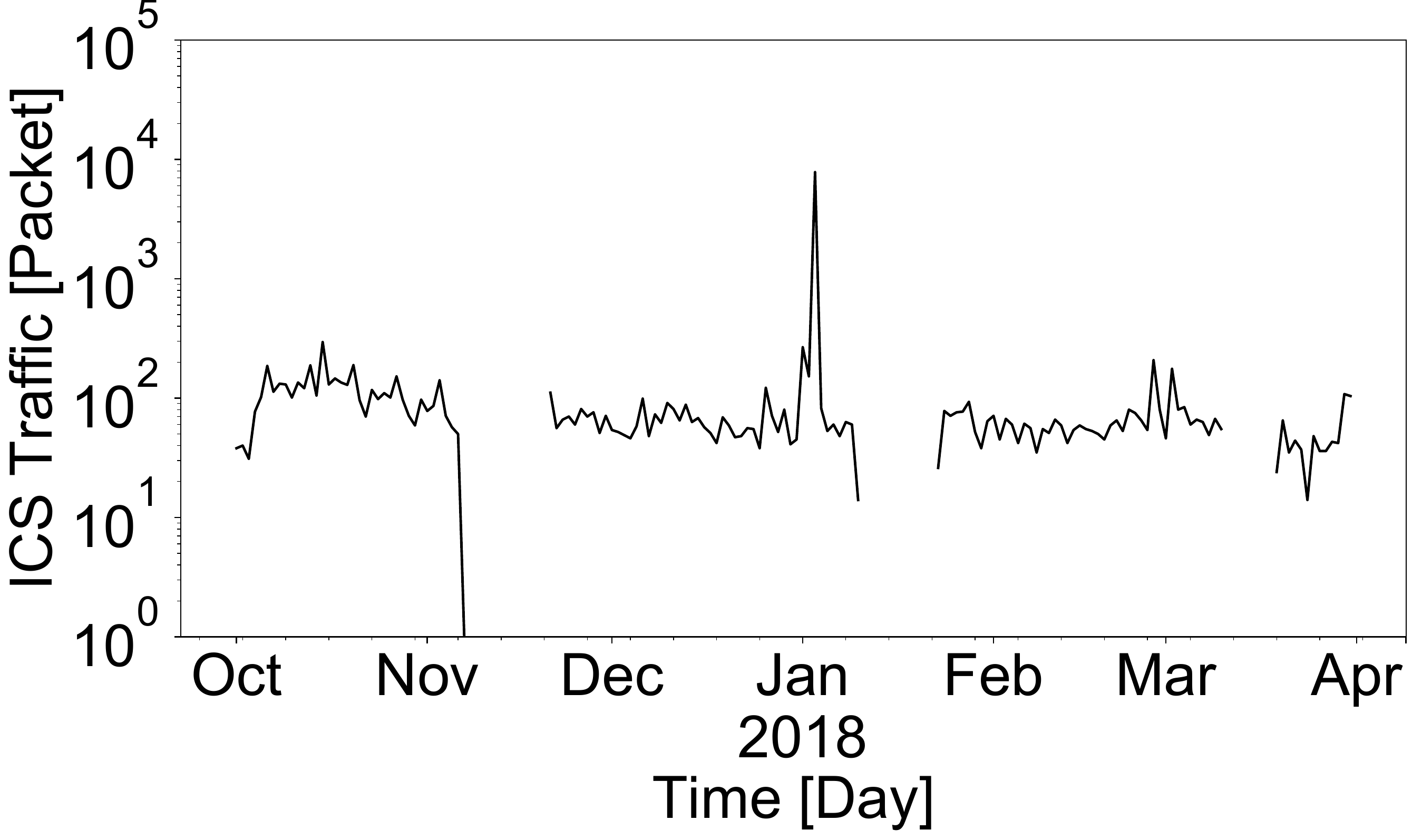}
    \caption{Internet Exchange Point (IXP)}
  \end{subfigure}
  \begin{subfigure}{.32\textwidth}
    \centering\includegraphics[width=\linewidth]{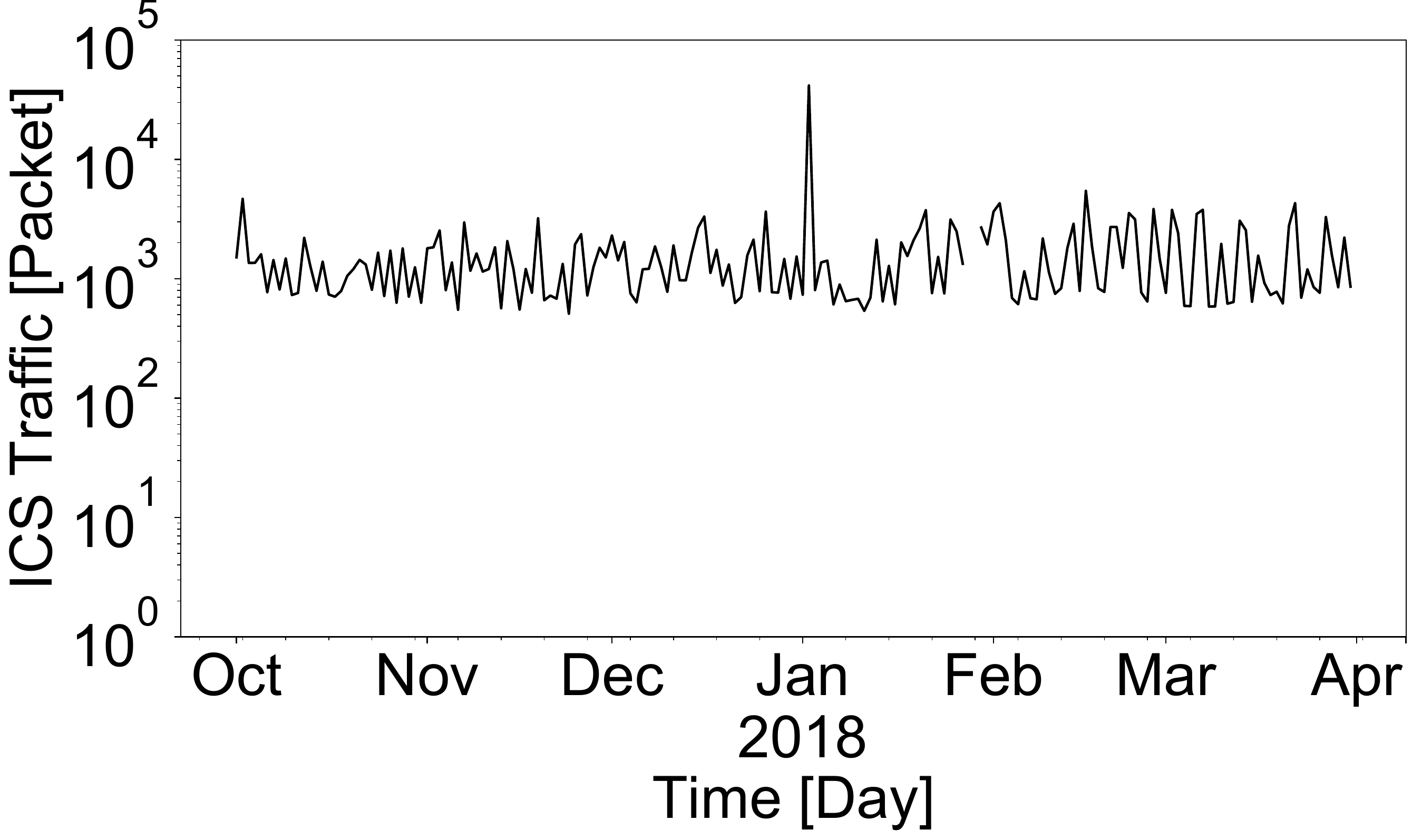}
    \caption{Internet Service Provider (ISP)}
  \end{subfigure}
  \begin{subfigure}{.32\textwidth}
    \centering\includegraphics[width=\linewidth]{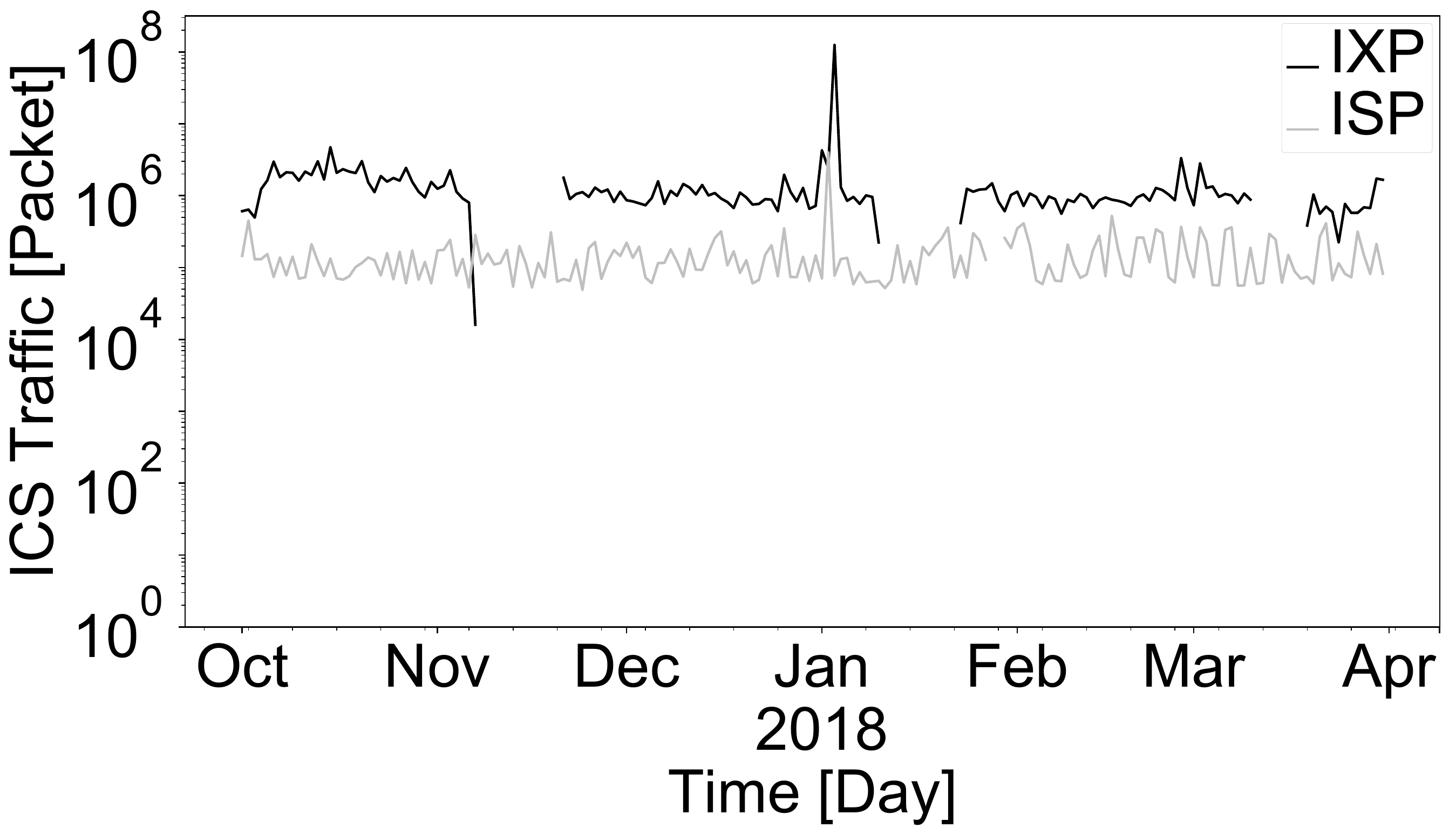}
    \caption{Extrapolation by sampling~rates.}
    \label{fig:packets_per_day_extrapolate}
  \end{subfigure}
  \caption{Number of inter-domain ICS packets per day at two different vantage points.}
  \label{fig:packets_per_day}
\end{figure*}

\section{Identification of ICS Traffic}
\label{sec:analysis-identification}

 Two challenges need to be tackled for analyzing inter-domain ICS traffic.
First, we need to reliably identify ICS traffic in global packet traces.
Second, we need to distinguish industrial (\ie transferred by real deployments) from non-industrial ICS traffic (\eg scanning) .
In this section, we propose our methodology to solve the first challenge, and tackle the second challenge in Section~\ref{sec:analysis-classification}.

\subsection{Collecting Traffic at Central Internet Vantage Points}

We passively collect traffic at two different Internet vantage points, an IXP and an ISP.
The two data source allow us to inspect traffic from two different perspectives, a rich interconnection fabric and an upstream provider.

Internet exchange Points (IXP) are centralized network infrastructures where heterogeneous domains intertwine.
We receive data from a large, regional IXP from Europe with over 100 member networks with a daily traffic peak of 560 GBit/s.
Due to the large traffic volume, flow data is not fully recorded but selectively sampled.
We analyse non-anonymized packets collected from October 2017 until April 2018 with a sample rate of $ \sim 2^{14} $.
The sampled packets are truncated after 128 bytes.
Flows from an IXP are inherently inter-domain.

Our second data source is the \emph{Measurement and Analysis on the WIDE Internet} (MAWI) archive ~\cite{MAWIData}.
This archive contains daily traces describing 15~minutes of full traffic captures from a transpacific Internet link between Japan and the United States.
We received a private MAWI data set with non-anonymized IP addresses and payload (96 bytes) for the same time range.

Non-anonymised flows allow for mapping with additional meta data, such as autonomous systems.
Please note that we are not allowed to release our data due to privacy constraints.

\subsection{Identifying ICS Traffic Candidates}

We explicitly do not want to implement new traffic classifiers as this conflicts with maintainability and reproducibility on the long-term.
Instead, we want to leverage existing tools.
We use Wireshark dissectors to find ICS traffic candidates.
Half of the ICS protocols can be dissected by Wireshark, as shown in \autoref{tab:ics-overview}.
Wireshark distinguishes between normal and heuristic dissectors (ND, HD).
Normal dissectors identify protocols based on well-known port numbers and check whether the packets comply with simple sanity checks.
If they fail, they forward the data to heuristic dissectors which apply pattern matching on protocol fields.

To verify the correctness of the Wireshark dissectors, we apply them on public ground truth data~\cite{icsPCAPS} and manually inspect the dissection of packet headers.
All dissectors except one work accurately and map operation codes to protocol actions, such as \textit{read} or \textit{write}.

Packet sampling at our vantage points does not store complete packets but only a pre-configured fixed size of the overall packet.
This limitation can lead to inaccuracies in identifying the application layer protocol because parts of the corresponding headers are missing.
For each protocol, we reduce the packet length of the ground-truth data byte-wise and detect the minimal packet length   required to identify the protocol correctly.
All but one protocol dissector require less than 96~bytes, see \autoref{tab:ics-overview}.
Considering that packets are truncated after 128~bytes at the IXP and 96~Bytes at the ISP, we can identify the ICS traffic candidates reliably.

\subsection{Sanitizing ICS Traffic Candidates}
\label{sec:sanitizing}

\begin{table}[t]
  \centering
    
  \caption{Effects of data sanitization process and the ratio of remaining ICS packets by vantage point.
    \label{tab:DataSanitization}}
  
  \begin{tabular}{l r r}

  \toprule
  & \multicolumn{2}{c}{Remaining Packets} \\
 \cmidrule{2-3}
& IXP 	& ISP  	\\ \midrule
\textbf{Sanitizing steps after Wireshark ICS detection} & 100\% & 100\%\\
\quad \onec\, Removal of tunnel packets 					& 99\%	& 99\%	\\
\quad \twoc\, Removal of malformed packets 			& 15\%	& 52\%	\\
\quad \threec\, Removal of NDPI fingerprintable packets	& 14\%	& 51\%	\\ \midrule

\textbf{Comparison with vanilla approach} \\
\quad Port-based detection relative to Wireshark & 3950\%  & 1340\% 	\\
  \bottomrule

  \end{tabular}
\end{table}

We do not rely blindly on the Wireshark dissectors.
We perform three data sanitizing steps to improve data quality: 
\onec\, We remove tunnel traffic so that we only obtain plain end-to-end traffic. 
This step mainly excludes ICMP unreachable messages, which encapsulate the original UDP packets.
Such backscatter packets are misclassified by Wireshark as ICS traffic. 
\twoc\,~We remove packets which Wireshark marks as \textit{malformed} or cases in which the dissector reports an \textit{error}.
This occurs when the protocol detection of a packet is successful, but the complete dissection fails due to header fields that do not comply with the protocol specification.
\threec\,~We cross-validate our data by applying NDPI~\cite{ndpiHomepage}, a leading open-source deep packet inspection software.
NDPI detects a broad range of protocols, but no ICS protocols.
We exclude every packet that NDPI is able to map to a known protocol since we consider such a packet to be a false positive. 

In \autoref{tab:DataSanitization}, we quantify the remaining packets after applying our sanitizing steps.
The data is shown relatively to the overall amount of identified ICS packets per vantage point.
85\% of the packets at the IXP are classified malformed, and 48\% at the ISP.
Wireshark detects ICS protocols although many header fields are set to unspecified values, such as unknown operation codes.
This highlights that Wireshark dissectors are rather optimistic and sanitizing is required for a reliable analysis.
The removal of packets identified by NDPI accounts for less than 1\%, which indicates a very low false-positive rate of our approach.
Finally, we compare our approach with a pure port-based detection.
Identifying ICS traffic only based on port numbers is not feasible as it leads to significant overestimation.

\section{Properties of ICS Traffic}
\label{sec:analysis-properties}

\subsection{Daily Patterns and Prevalence of Inter-Domain ICS Traffic}

During our measurement period, we identified 19k ICS packets at the IXP and 310k ICS packets at the ISP after sanitization.
\autoref{fig:packets_per_day} shows the number of daily ICS packets at the IXP and ISP.
For better comparison, we consider the different sampling intervals and extrapolate the values (see \autoref{fig:packets_per_day_extrapolate}).
The daily ICS traffic at the IXP and ISP is constant apart from one anomalous peak at each vantage point.
The traffic peak at the IXP is due to a large number of EthernetIP packets (217.5 MB/s traffic peak) during 10~minutes on January 3, 2018.
The destination is a single IP address and the traffic is sent from several sources located in two autonomous systems.
The traffic peak at the ISP consists of BACnet messages from 76~source IP addresses to 41,000~destination IP addresses.
This event took place one day before the IXP peak.
We observe uniformly distributed BACnet read messages, which indicates load balancing between scanning nodes.
All sources relate to Rapid7 Sonar, a company that performs regular Internet-wide BACnet scans.

\begin{figure}
\centering
  \centering
  \includegraphics[width=0.95\linewidth]{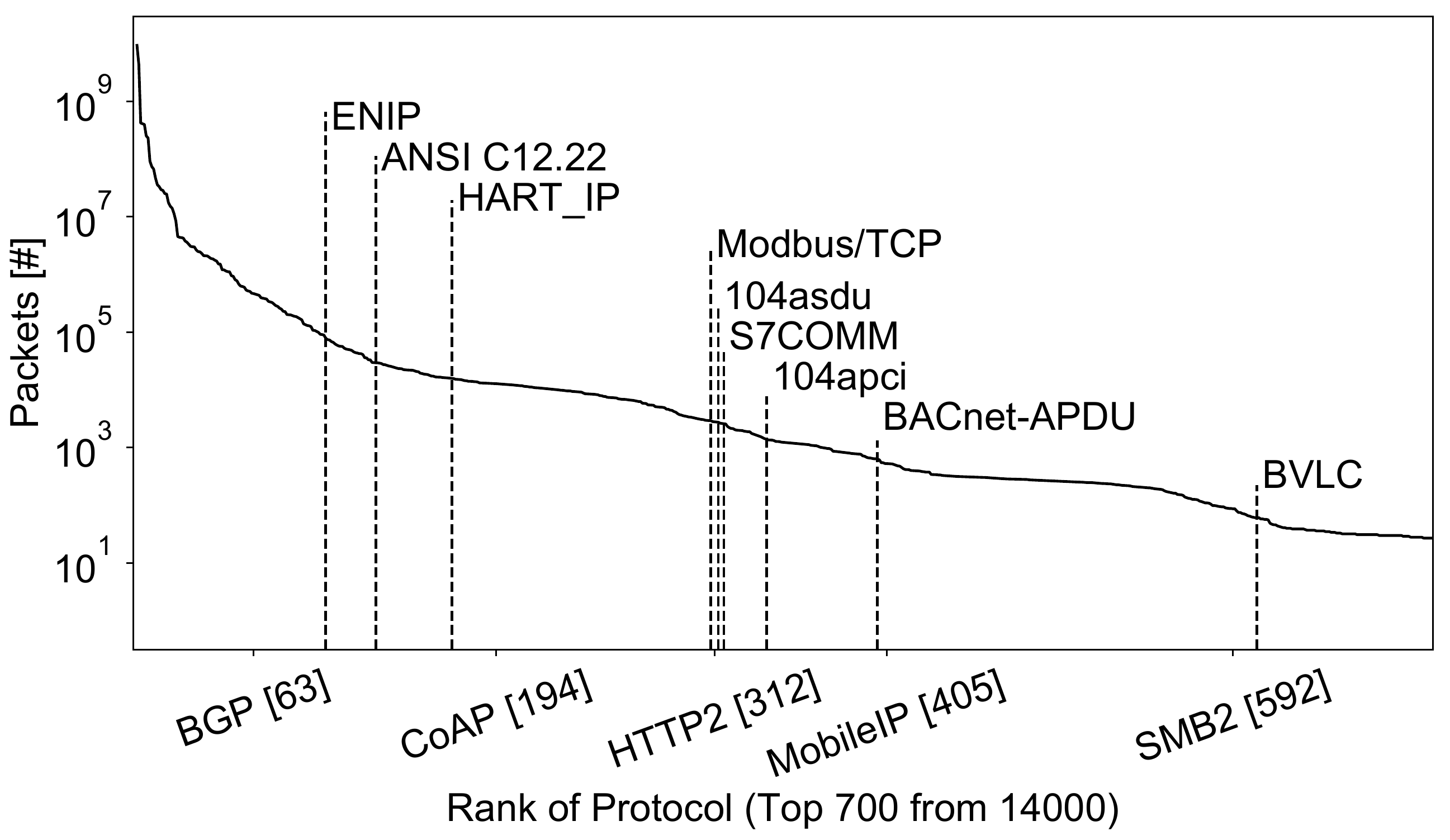}
  \caption{Protocols ranked by packet frequency as reported by Wireshark (non-sanitized), observed at a big national IXP during 6~months. ICS protocols are emphasized among some well-known protocols. Ranks are noted in brackets.}
  \label{fig:proto_rank}
\end{figure}

Compared to the total traffic volume, ICS inter-domain traffic is low.
ICS packets only account for $\approx$~0.0001\% of all sampled packets at the IXP and $\approx$~0.002\% at the ISP.
However, putting ICS traffic into perspective of well-known non-ICS protocols, ICS traffic is likewise prevalent, which we show in \autoref{fig:proto_rank}.
To allow for comparability, this graph visualizes the non-sanitized data set because implementing a sanitization process for non-ICS protocols would be out of scope of this work.
This result emphasizes that ICS traffic should not be neglected.

\subsection{ICS Message Types: Request vs. Reply}
\label{sec:messageType}

We refer to packets sent to a known ICS port as requests, and packets originating from a known ICS port as replies. 
Protocols with  balanced request-reply ratios are likely to be used in a legitimate way since ICS communication patterns follow a common client server scheme.
Observing significantly enhanced requests  may have two reasons:
\one heartbeats sent from sensors to central servers that do not confirm the reception; 
\two scan traffic that reaches hosts which do not offer the corresponding service.

We analyze the ratio of requests and replies per protocol in more detail, check left-hand side of \autoref{tab:correlation-overview}. 
We observe a tendency towards requests exceeding replies.
Only at the IXP, HartIP and C12.12 show a balanced request-reply ratio.
Strikingly, BACnet is very request-heavy across both vantage points.
This might be an indication for non-industrial ICS traffic, which we will investigate further in \autoref{sec:analysis-classification}. 

\subsection{ICS Traffic Sent to and Received from Autonomous Systems}
\label{sec:InterDomainView}

To better understand the ICS ecosystem from a networking perspective, we map each source and destination IP address of a sampled packet to autonomous system numbers (ASN).
We use daily data from the RIPE RIS project and topological information from the IXP for assigning ASNs.

\begin{figure}
  \center
  \includegraphics[width=\linewidth]{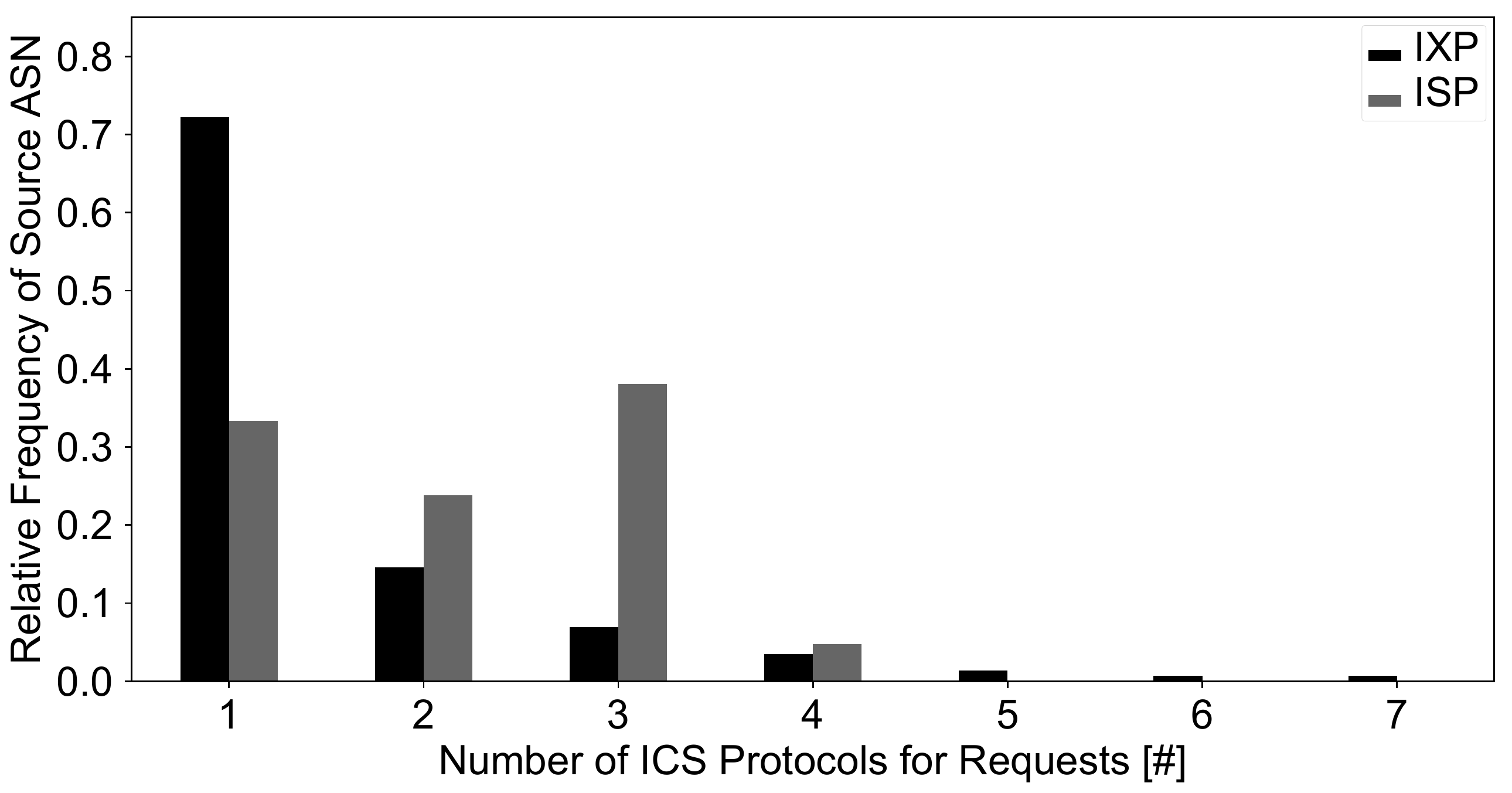}
  \caption{Number of ASes sending different ICS protocol requests. Since ICS deployments are rather specific deployment and
  bound to a single manufacturer, we rate several ICS protocols originating from a single AS as suspicious.}
  \label{fig:ICSinASNRequest}
\end{figure}

Autonomous systems (AS) which are the origin of request traffic via multiple ICS protocols host either scanners or heterogeneous ICS monitoring services.
In our sanitized data sets, more than 70\% of the ASes host nodes that deploy a single ICS protocol, see \autoref{fig:ICSinASNRequest}.
We find four cases of ASes creating requests for \textgreater 4 distinct ICS protocols.
Three are eyeball providers and one is a webhoster.
These types of networks are common to connect scanners, which we detect in \autoref{sec:analysis-classification}.

\begin{table}[t]
  \centering
    
      \caption{Amount of successful reverse DNS lookups of source IP addresses per scan project.
    \label{tab:ScannerDNSLookup}}
  
  \begin{tabular}{l r r}

  \toprule
					& IXP		& ISP	\\ \midrule
\# Unique source IP addresses	& 1504 		& 223	\\ \midrule
\# Resolvable Censys IP addresses			& 105		&	n/a	\\	
\# Resolvable Rapid7Labs IP addresses		& 	7		&	56	\\
\# Resolvable Kudelski Sec. IP addresses    & 	0		&	0 \\
\# Resolvable Shodan IP addresses			& 	23		&	25	\\ 

  \bottomrule

  \end{tabular}
\end{table}

\begin{table*}[t]
  \centering
    
  \caption{Relative amount of industrial ICS traffic after applying different filter rules on the observed ICS traffic.
    \label{tab:correlation-overview}}
  
  \begin{tabular}{l r r r r r r r r r r r r }

  \toprule
     & \multicolumn{4}{c}{Request response ratio} & \multicolumn{8}{c}{Traffic share after applying filters} 
	 \\
	 \cmidrule(rl){2-5} \cmidrule(rl){6-13}
     & \multicolumn{2}{c}{\# ICS Packets} & \multicolumn{2}{c}{Share of Requests} & \multicolumn{2}{c}{Excluding scanners} & \multicolumn{4}{c}{Excluding captured honeypot data} & \multicolumn{2}{c}{Excluding both}
     \\
	 \cmidrule(rl){2-3} \cmidrule(rl){4-5} \cmidrule(rl){6-7} \cmidrule(rl){8-11} \cmidrule(rl){12-13}
     & IXP & ISP & IXP & ISP & IXP & ISP & \multicolumn{2}{c}{IXP} & \multicolumn{2}{c}{ISP} & IXP & ISP
     \\
	 \cmidrule(rl){8-9} \cmidrule(rl){10-11} 
			& & & & & & & HP$_{ICS}$ & HP$_{all}$	& HP$_{ICS}$  &  HP$_{all}$ \\ \midrule
Total & 19,060 & 310,996 & 81\% & 99\% & 97\% & 46\% & 97\% & 96\% & 15\% & 1.5\% & 96\% & 1.5\%\\ \midrule
BACnet      	 & 568 &  89,922 & 98\% & 100\%					 & 15\% & 7\% & 25\% & 11\%  		& 40\% & 1\% & 10\% & 1\%	\\
C12.22         	 & 1,559 & 24 & 63\% & 29\%				 & 100\% & 100\% & 100\%	& 99\%    	& 100\% & 100\% & 99\% & 100\%	\\
DNP3         	 & 2 &  2,424 & 100\% &	100\%			 & 100\% & 99\% & 0\%	& 0\%   	& 0.4\%	& 0.1\%	 & 0\% & 0\% \\
EthernetIP   	 & 9,171 & 171,804 & 94\% & 99\% 				 & 99\% & 75\% & 98\%	& 98\% 	& 5\%	& 0.02\% & 98\% & 0\%\\
HartIP      	 & 126 & 46,783 & 62\% & 92\%				 & 65\% & 9\% & 62\%	& 62\%  	& 9\% 	& 8\% & 62\% & 8\%\\
IEC60870      	 & 2,511 & 13 & 13\% & 38\%				 & 100\% & 100\% & 100\%	& 99\%     	& 100\%	& 100\%	 & 99\% & 100\%\\
Modbus			 & 2,547 & -- & 95\% & --		 & 100\% & -- & 100\%	& 100\%			& --	& -- & 100\% & --\\
Siemens			 & 2,576 & -- & 99\% & --			 & 100\% & -- & 100\%	& 100\%			& --	& -- & 100\% & --\\
  \bottomrule

  \end{tabular}

\end{table*}

\section{Identification of Industrial and Non-industrial ICS Traffic}
\label{sec:analysis-classification}

Separating non-industrial from industrial ICS traffic allows us to identify the vulnerable part of the ICS ecosystem more precisely.
We classify ICS traffic at our vantage points as non-industrial if the captured IP addresses belong to scan projects or have been observed at honeypots, as those indicate non-ICS hosts.

\subsection{Filter Traffic of Common Scan Projects}
\label{sec:ScannerNames}

Several projects scan for ICS hosts on a regular basis and thus contribute to non-industrial inter-domain ICS traffic.
The most common projects are Censys, Shodan, Rapid7, and Kudelski. 
Censys, Rapid7, and Kudelski publicly document the IP prefixes from which they initiate scans.
We use these prefix lists to identify scanners by marking an observed source IP address as scanner if the source IP address is covered by one of the prefixes.

To identify scanners that are not part of the documented IP prefixes, we perform reverse DNS lookups on all source IP addresses captured at our vantage points.
By reviewing the assigned names manually, we find Censys, Rapid7, and Shodan scanners (\eg \emph{pirate.census.shodan.io} and \emph{scanner2.labs.rapid7.com}).
Note that we cannot identify any names that relate to Censys at the ISP because Censys performs scans between $\approx$ 8:00am and $\approx$~6:00pm (UTC), whereas the ISP dumps include 15~minutes packet captures starting at 5:00am~(UTC).

\autoref{tab:ScannerDNSLookup} shows the amount of successful reverse DNS lookups. The IXP and ISP share 86~source IP addresses, predominantly Shodan and Rapid7 scanners.
The 5~most common source IP addresses at the ISP resolve to Shodan names and are located in Quasi Networks, an autonomous system which is also well-known for hosting malicious nodes~\cite{QuasiNetworksBlog}.

\begin{figure*}
  \center
  \begin{subfigure}{.49\textwidth}
    \centering\includegraphics[width=\linewidth]{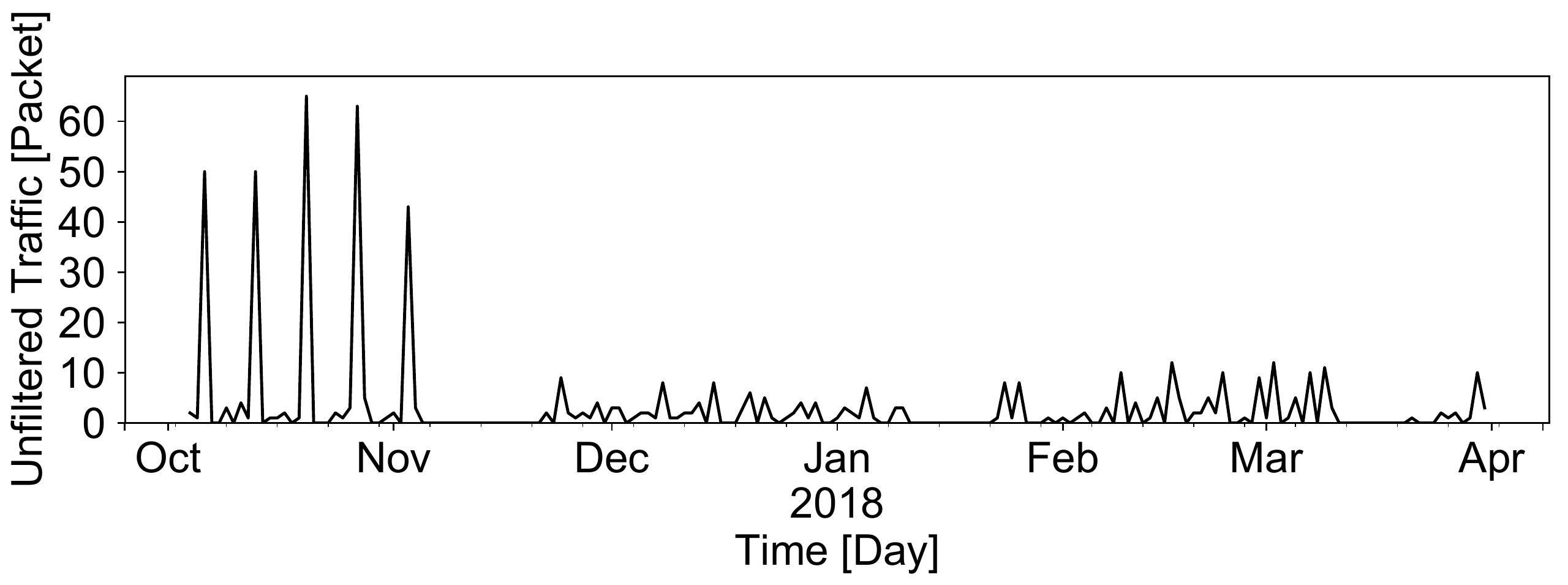}
    \caption{IXP -- Total BACnet Traffic.}
  \end{subfigure}
  \hfill
  \begin{subfigure}{.49\textwidth}
    \centering\includegraphics[width=\linewidth]{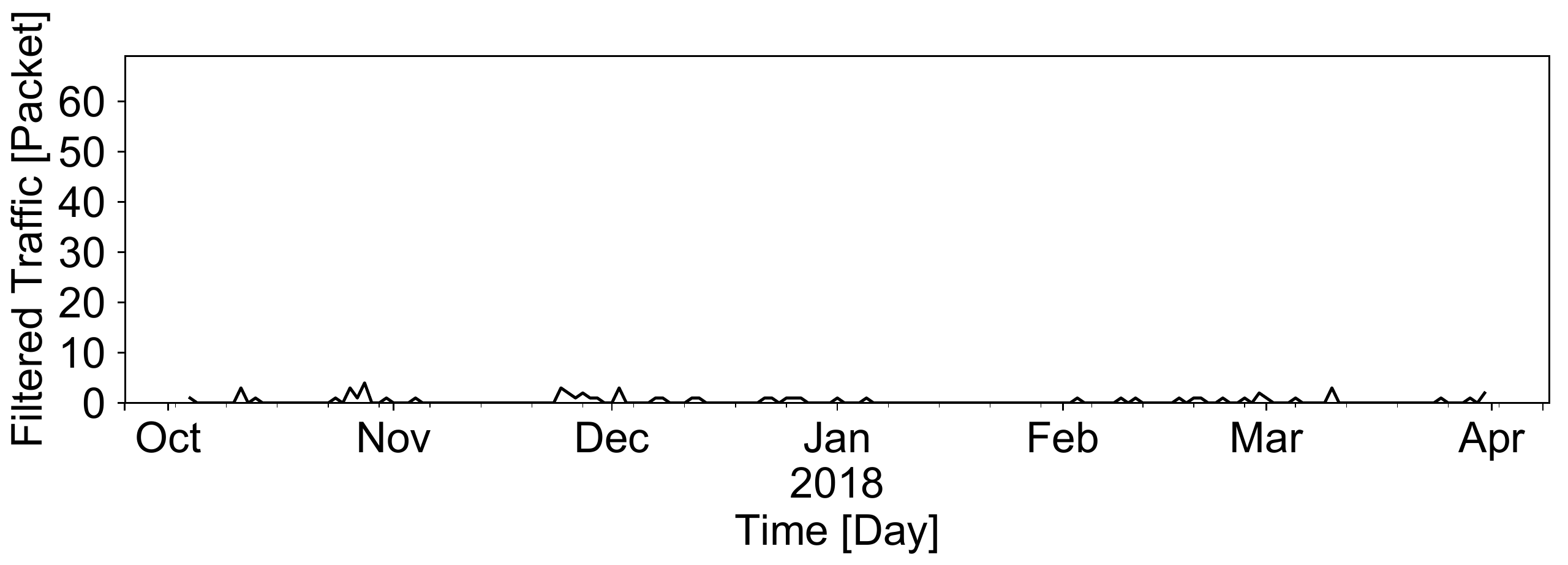}
    \caption{IXP -- Industrial BACnet Traffic.}
  \end{subfigure}

  \begin{subfigure}{.49\textwidth}
    \centering\includegraphics[width=\linewidth]{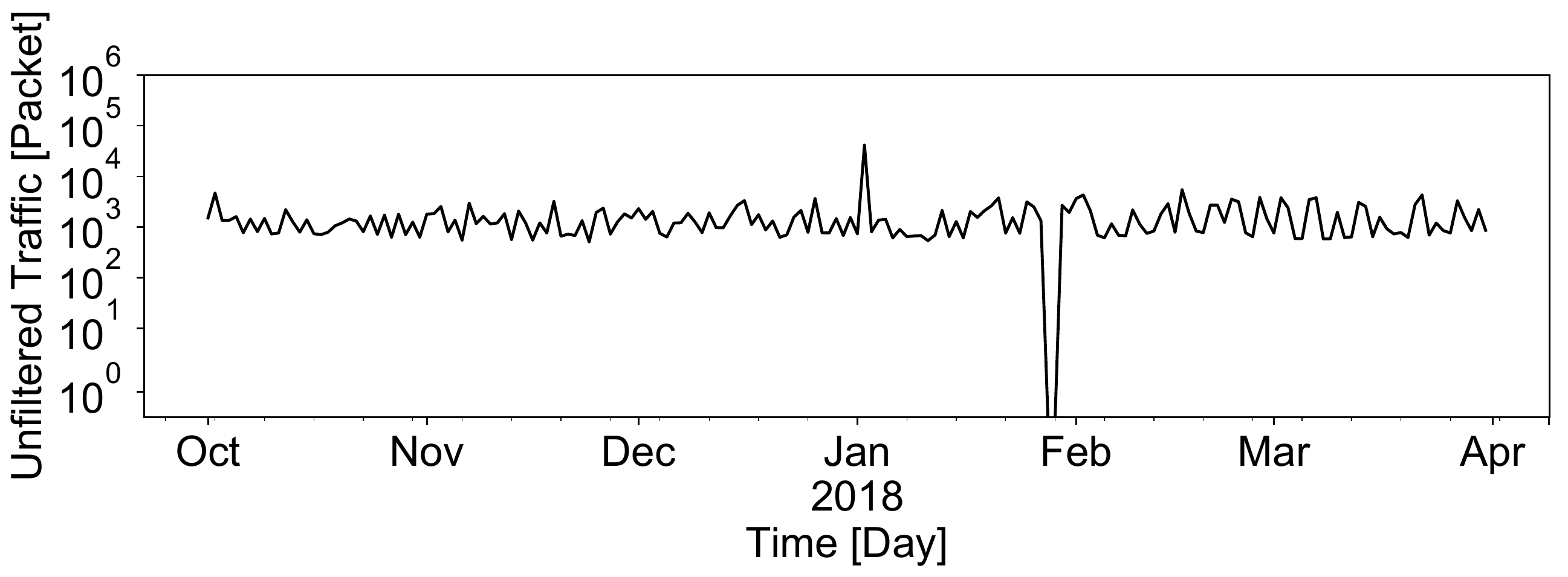}
    \caption{ISP -- Total ICS Traffic.}
  \end{subfigure} 
  \hfill
  \begin{subfigure}{.49\textwidth}
    \centering\includegraphics[width=\linewidth]{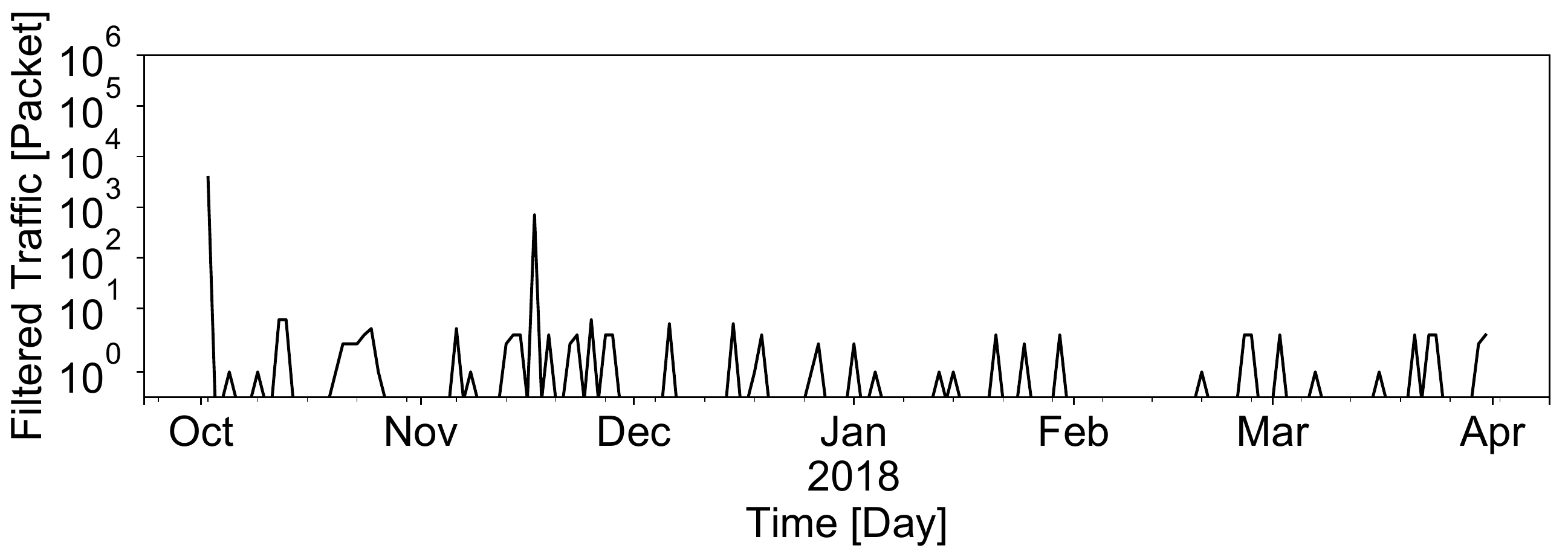}
    \caption{ISP -- Industrial ICS Traffic.}
  \end{subfigure}  
  \caption{Daily amount of all ICS traffic versus industrial ICS traffic visible at the IXP and ISP.}
  \label{fig:TrafficFilter}
\end{figure*}

\subsection{Filter Traffic of Other Non-ICS Hosts}
\label{sec:Honeypot}

To account for other hosts that create non-industrial ICS traffic (\eg attackers), we leverage data from honeypots.
Conpot \cite{ConpotHP} is the de-facto standard ICS honeypot but supports only five ICS protocols, one currently under development. 
Conpot implements limited variances in responses, which makes it easy to unmask as a honeypot.
Thus, we argue to utilise transport layer honeypots in order to measure a broad scope of activities on ICS ports. 

We deploy Honeytrap \cite{HoneytrapHP} in \one a university network and \two a darknet, a network not offering any public services.
Based on these honeypots, we identify suspicious source IP addresses.
We create two lists: HP$_{all}$, which stores all IP addresses observed at the honeypots, and a subset of this list, HP$_{ICS}$, which stores IP addresses that sent requests to at least one ICS port.
HP$_{all}$ consists of 244k IP addresses and HP$_{ICS}$ of only 3700 IP addresses (1.5\%) from 619 ASes. It is worth noting that our honeypots also capture sources of the well-known scan projects. 
224~IP addresses in HP$_{all}$ are from Censys scanners.

We now correlate ICS traffic from our vantage points with the honeypot data.
For every observed ICS packet,  we check whether the source or destination IP address is present in HP$_{all}$ or HP$_{ICS}$, see \autoref{tab:correlation-overview}.

At the IXP, the overlap is minimal, which means that a significant amount of industrial ICS traffic is visible.
96\% of ICS traffic is industrial based on HP$_{all}$, 97\% based on HP$_{ICS}$.
We perform a comparison per protocol and correlate 506~BACnet packets with HP$_{all}$, which represent 89\% of the total BACnet packets at the IXP.
These packets are classified as non-industrial ICS traffic and filtered.
The results are stable, even if we only consider HP$_{ICS}$.

At the ISP, less industrial ICS traffic is visible.
Filtering by HP$_{all}$, we find only 1.5\% of the traffic to be industrial.
However, the filtering is less effective if we only consider HP$_{ICS}$, especially for BACnet.
The results indicate that it is beneficial to include honeypot information from non-ICS ports.

\subsection{Benefits of Combining Filter Rules}
\label{sec:combiningFilterRules}

To summarize the results from our previous filter steps, we provide an overview of the impact of the different filters.
\autoref{tab:correlation-overview} shows the relative amount of ICS traffic that remains when traffic from scanners (identified by DNS names and IP prefixes), honeypots, or both is excluded.

While we classify 96\% of the traffic at the IXP as industrial, we see only 1.5\% of industrial traffic at the ISP.
Interestingly, more than half of the traffic at the ISP can already be classified as non-industrial only by excluding public scanners, \ie without maintaining a dedicated infrastructure such as honeypots.
However, even though maintaining a honeypot introduces additional complexity, its data is necessary to provide a more complete view on distinguishing industrial and non-industrial traffic.

ICS protocols show similar trends for the the share of non-industrial traffic across both vantage points.
The substantial difference for EthernetIP is caused by a Shodan scan of a complete prefix range at the ISP.

We show the potential effects of filtering non-industrial ICS traffic over six months in \autoref{fig:TrafficFilter}.
This enables us to describe the impact of non-industrial traffic over time.
At the IXP we focus on BACnet as this protocol is severely affected by non-industrial activity.
We make two observations:
\one At the IXP, non-industrial traffic consists mainly of ephemeral spikes at the beginning of our measurement period.
\two At the ISP, the non-industrial traffic shows a very constant daily activity.
After filtering at both vantage points, we obtain only a few industrial ICS packets per day which allows even for manual inspection of the ICS traffic.

\section{Properties of ICS Industrial and Non-Industrial Traffic}
\label{sec:trafficFeatures}

\begin{table}[t]
  \centering
  \caption{Successful transport and application layer handshakes during Censys scans. 
    \label{tab:ScannerHandshakes}}  
  \begin{tabular}{l r r l}
  \toprule
      			& \multicolumn{2}{c}{\# ICS hosts detected by Censys} \\ \cmidrule{2-3}
Protocol 		& Transport Scan		& Application Scan 	\\ \midrule
BACnet			& 31,735                 & 31,154 & (98\%) \\
Modbus			& 8,400,058               & 126,984 & (2\%) \\
Siemens	S7		& 7,202,828               & 24,946 & (0.5\%) \\
  \bottomrule
  \end{tabular}
\end{table}

\begin{table}
  \centering
      \caption{Ratio of ICS hosts observed at the IXP and Censys.
    \label{tab:ScannerIXPhosts}}
  \begin{tabular}{l r r}
  \toprule
      & \multicolumn{2}{c}{\% ICS hosts that overlap with Censys} 	\\ \cmidrule{2-3}
Host Type at IXP 		& Transport Scan	& Application Scan	\\ \midrule
BACnet Source			& 0\%	            & 0\%				\\
BACnet Destination 		& 0\%				& 0\%				\\
Modbus Source			& 3\%	            & 0\%				\\
Modbus Destination 		& 35\%				& 0\%				\\
Siemens Source  		& 0\%	            & 0\%	            \\
Siemens Destination 	& 65\%				& 65\% 				\\
  \bottomrule
  \end{tabular}
\end{table}

\begin{figure*}
\centering
\begin{minipage}{0.37\textwidth}
\centering
\includegraphics[width=0.9\linewidth]{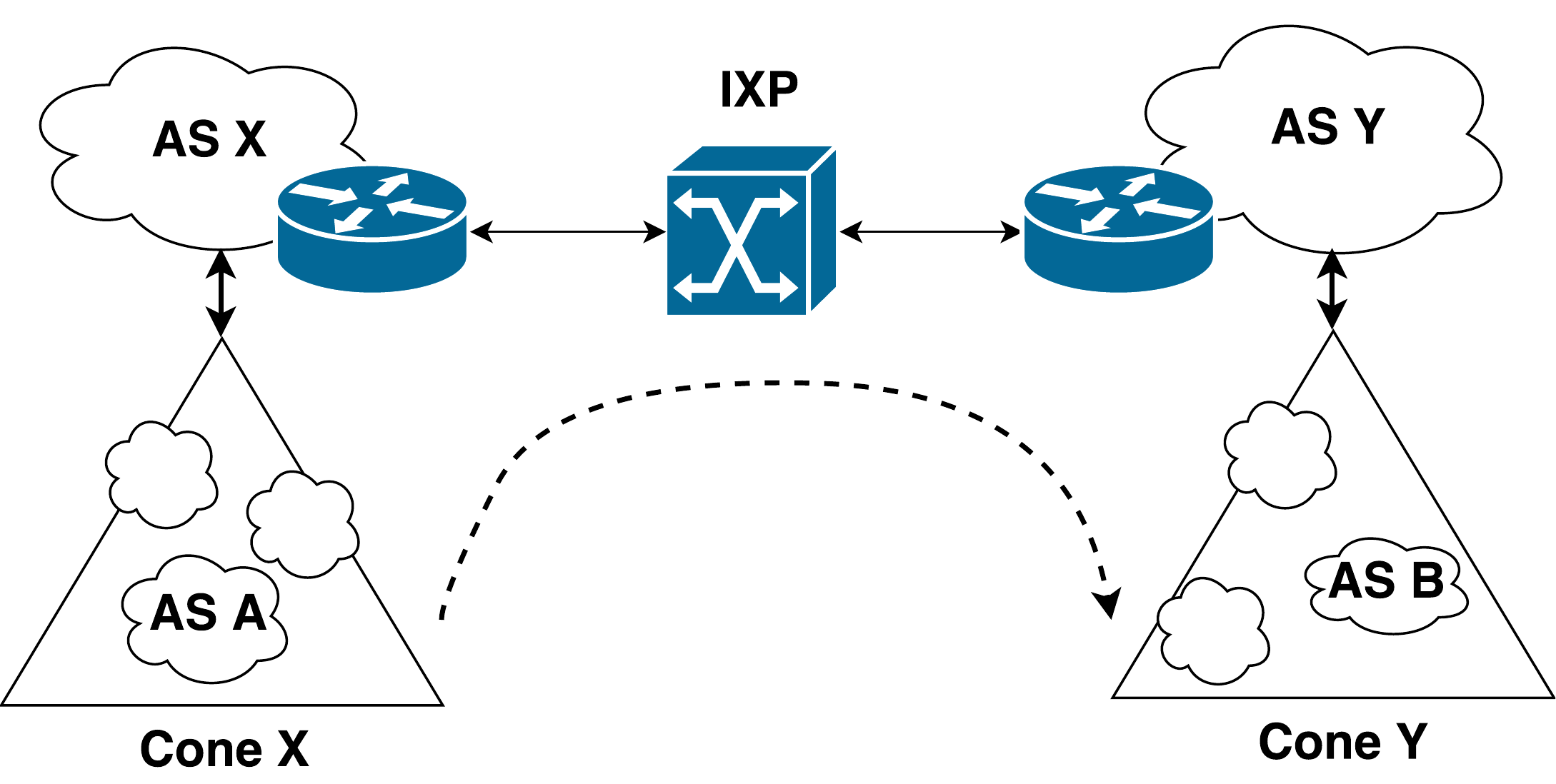}
\captionof{figure}{Example of cone to cone communication with ingress AS X and egress AS Y.}
\label{fig:IXPtopology}
\end{minipage}
\hfill
\begin{minipage}{0.6\textwidth}
\centering
\setcounter{table}{7} 
\footnotesize\captionof{table}{Relative ratio of traffic transitions for three ICS protocols at IXP.
Non-industrial traffic originates exclusively from cones and thus is not local.}
\label{tab:TrafficLocalityAfterScrubbing}
  \begin{tabular}{l r r r r r r r}
  \toprule	
	& \multicolumn{3}{c}{Industrial} & \multicolumn{3}{c}{Non-Industrial} 
\\
\cmidrule(rl){2-4}
\cmidrule(rl){5-7}
				& BACnet &	HartIP &  EthernetIP & BACnet &	HartIP &  EthernetIP \\ \midrule
Member to Member 	& 30\% & 22\% & 5\% & 0\% & 0\% & 0\%
\\
Member to Cone 		& 24\% & 51\% & 29\% & 0\%  & 0\% & 0\%
\\
Cone to Member		& 19\% & 9\% & 6\% & 46\% & 79\% & 52\%
\\
Cone to Cone 	& 27\% & 18\% & 60\% & 54\% & 21\% & 48\%
\\
\midrule
\# Flows & 59 & 78 & 9006 & 509 & 48 & 165
\\
\bottomrule
\end{tabular}
\end{minipage}
\end{figure*}

\subsection{Detecting ICS Hosts Protected by Firewalls}
\label{sec:combiningFilters}

ICS devices might be protected by firewalls which grant access only from specific hosts.
We analyze this by comparing IP addresses observed in our passive data with IP addresses of ICS devices revealed by active scans.
To reduce overhead on the Internet infrastructure~\cite{klwr-tbicr-16b}, we do not implement our own active probing but use data from Censys.
Censys continuously scans the entire public IPv4 address space fast~\cite{dwh-zfiws-13,7906943}, implements full transport  and application layer handshakes~\cite{dwh-zfiws-13}, and releases weekly snapshots.
We compare 3~ICS protocols for which we found industrial traffic and which are scanned by Censys during our measurement period: Siemens S7, Modbus, and BACnet.

First, we check how many ICS hosts are detected by Censys on the transport and application layer (see \autoref{tab:ScannerHandshakes}).
Despite many successful transport layer handshakes, Modbus and S7 exhibit a very low success rate on the application layer.
We argue that this is related to the use of lower port numbers that are more likely to be used by other applications which listen on the corresponding port.
This complies with our previous results which showed that port-based ICS detection is misleading (see \autoref{sec:sanitizing}).

Now, we compare with ICS hosts observed at our vantage points.
We compute the fraction of source or destination IP addresses that have been discovered by Censys (see \autoref{tab:ScannerIXPhosts}) and for which we see communication in our passive data, \ie completely unprotected nodes.
At the IXP, 35\% of the Modbus and 65\% of Siemens destinations are already known because of the transport layer scan.
At the ISP, we do not find any correlation, \ie none of the ICS devices that are visible in our ISP traffic data set have been captured by active scans.
This is very likely due to port-based access control lists which only allow communication between pre-configured hosts.

We find 3~source IP~addresses that respond to Modbus transport layer scans but do not establish successful application layer sessions based on Censys.
However, based on our traffic traces, each of these hosts has sent about 45~Modbus packets.
One host is sending packets to a solar energy consulting agency.
These results indicate cases of secure ICS services but unprotected ICS traffic.

\subsection{Host Stability of Industrial ICS Traffic}

Host stability describes how often a host is visible at our vantage points with respect to an activity span.
For each destination IP address in the industrial ICS traffic, we calculate the size of the activity window $w$ in days
(\ie time-lag between first and last day of occurrence) and the number of active days $n$ within this time window. 

We assume that as soon as an ICS network is in place an embedded ICS device and an ICS control station will frequently exchange ICS traffic.
Furthermore, we assume static assignment of IP addresses to those devices as this will ease operational maintenance (\eg configuration of firewall rules).
Following both assumptions, hosts will achieve high host stability in case of real ICS networks, \ie the same IP address will appear for several days.

The IXP and ISP results differ significantly.
In the IXP data set, the most stable host communicates almost every day ($w=179, n=146$).
In contrast to this, in the ISP data set, hosts communicate less than 4\%, relatively to the overall activity~span.

To better understand whether stable hosts belong to a real ICS deployment, we map IP addresses to additional meta data: reverse DNS records and \texttt{whois} data.
Based on this, we find that hosts are operated by a building company (\emph{max-boegl.de}; $w=179, n=146$), a trade and transport company (\emph{Handel Uslugi Transport Ewa Cielica}; $w=159, n=98$), and a industrial service and consulting company in the field of solar energy (\emph{enerparc.com}; $w=90, n=36$).
The high number of active days, despite the sampling, indicates a high exchange of messages.
Interestingly, these hosts are not marked as ICS hosts by Censys, indicating the role of an ICS monitoring station.
In the data set of our transnational ISP, we do not find evidence for ICS companies. 

\setcounter{table}{6} 
\begin{table}
      \caption{Relative ratio of domestic traffic for three ICS protocols, compared to the overall traffic of each protocol at the IXP.
    \label{tab:DomesticTrafficAfterScrubbing}}
  \begin{tabular}{l c c c c c c }
  \toprule
	& \multicolumn{3}{c}{Industrial} & \multicolumn{3}{c}{Non-Industrial} 
\\
\cmidrule(rl){2-4}
\cmidrule(rl){5-7}
				& BACnet &	HartIP &  EthernetIP & BACnet &	HartIP &  EthernetIP \\ \midrule
  & 29\% & 24\% & 1\% & 0\% & 0\% & 0.5\% \\
  \bottomrule
  \end{tabular}
\end{table}

\subsection{Locality of Non-Industrial Traffic}

We analyze the locality of industrial and non-industrial ICS traffic.
Less local traffic is more likely to be part of Internet-wide scanning activities, whereas some ICS stakeholders may consider locality as reason not to protect (industrial) ICS traffic.
We distinguish between topological and geographical locality.

\autoref{fig:IXPtopology} shows a typical inter-domain topology at an IXP.
In addition to a source and destination AS, packets may traverse \emph{ingress} and \emph{egress ASes} directly connected at the IXP.
ASes which send or receive packets over an IXP member are in the cone of this member.
We refer to traffic as IXP local, if the following condition applies: $Source\, AS==Ingress\,AS \quad \land \quad Egress\, AS==Destination\,AS$.
From a topological point of view, IXP local traffic is more \textit{trustworthy}, because both ASes peer directly with each other (maybe via a route server).
In contrast to this, communication from cones is rather expected from Internet-wide scanners, which are located in edge networks.
At the IXP, 
90\% of the BACnet and 40\% of the HartIP traffic is non-industrial.
Comparing peering transitions for these two protocols with EthernetIP, which exhibits only 2\% non-industrial traffic, shows a clear distinction, see Table~\ref{tab:TrafficLocalityAfterScrubbing}.
Non-industrial traffic originates only from the cones of the IXP-members, hence is not local at the~IXP.

Assuming that critical infrastructures are scanned by malicious hosts and proxies from ASes located in foreign countries, we also check how often traffic is locally bound to a country.
We do this by mapping the IP addresses to country codes based on MaxMind~\cite{pukdg-gdu-11}.
If the source and destination IP addresses are located in the same country, we call the traffic \emph{domestic}.
\autoref{tab:DomesticTrafficAfterScrubbing} presents the results of our analysis of domestic traffic.
Although industrial traffic is also exchanged across country borders (which might happen in the case of, \eg global transport companies), there is a clear trend for non-industrial traffic:
Non-industrial traffic is strictly non-domestic, which highlights globally distributed scanning activities.
On the other hand, up to 29\% of the industrial traffic is local, which makes it easy to contact and train the ICS network operators in charge.

\section{Conclusion}
\label{sec:conclusion}

In this paper, we analyzed the unprotected traffic of protocols that interconnect industrial control systems (ICS).
Our key results obtained from an IXP and an ISP perspective, \ie the Internet Core, read: 

\paragraph{ICS traffic identification is painful}
Common open source tools for traffic classification and analysis do not identify ICN traffic reliably. 
Due to the limited deployment of ICS protocols, there is a lack of fingerprinting tools.
We introduced and explored an advanced but lean approach to detect ICS protocols.
Our methodology is based on common Wireshark dissectors, but introduced several sanitizing steps that reduce the number of false positives.
Given that we have identified ICS scanners as well as industrial ICS deployments in our traffic traces, we are confident with our true positives.

\paragraph{Unprotected ICS traffic is  visible at the IXP}
After sanitizing our data, we found over 330k ICS packets and one anomalous traffic peak at each vantage point.
As Internet-wide ICS scanners operate since several years, it comes as no surprise that inter-domain ICS traffic exists.
Hence, we developed a classification mechanism to differentiate between industrial and non-industrial ICS traffic.
The  96\%-share of unprotected industrial ICS traffic at the IXP is alarming.
Since we observe a regional IXP, cooperating companies from the same region might exchange ICS traffic.
In contrast, our ISP data represents a transnational link between USA and Japan, representing the bridge between geographically distributed transit networks.
Intuition suggests, that distributed ICSs are rather local in deployment.
Our results confirm this intuition. With only 1.5\% industrial ICS traffic, the ISP is mainly confined to scans.

\paragraph{New, stable ICS deployments detected}
We isolated (non-) industrial ICS traffic, and could classify  ICS packets that were exchanged by hosts such as known from the Censys scan project.
We also discovered previously undetected ICS devices, though, that belong to real ICS eco-systems.
We identified cases of very stable hosts, \ie hosts that exchange ICS traffic regularly.
Such hosts are vulnerable to traffic manipulation attacks on a daily basis.
We spotted topological features for non-industrial ICS traffic.
Such traffic originates at IXP-cones and is not domestic, \ie source and destination are not located in the same country.

\paragraph{Raising awareness of potential ICS attacks}
The insights of this paper help to find unprotected ICS traffic and inform responsible stakeholders for improving protection.
They also allow to deploy a long-term monitoring system that can observe malicious inter-domain ICS activities.
Solutions already exist (such as SSH tunnel, VPN, \etc), but are not yet deployed, leaving ICS data exposed to eavesdropping and traffic manipulation attacks.

\clearpage
\balance

\bibliographystyle{IEEEtran}
\bibliography{bib/bibliography}

\end{document}